\documentclass[letterpaper,twocolumn,10pt]{article}
\usepackage{usenix2019_v3}

\usepackage{tikz}
\usepackage{amsmath}
\usepackage{booktabs} 
\usepackage{hyperref}
\usepackage{graphicx}
\usepackage{algorithm}
\usepackage{float}
\usepackage{subcaption}
\usepackage{titlesec}
\usepackage[noend]{algpseudocode}
\usepackage{nopageno}

\titleformat*{\subsubsection}{\large\bfseries}

\begin{document}


\date{}

\title{\Large \bf WaveGuard: Understanding and Mitigating Audio Adversarial Examples}

\author{\text{*}Shehzeen Hussain, \text{*}Paarth Neekhara, Shlomo Dubnov, Julian McAuley, Farinaz Koushanfar\\
University of California San Diego\\
{\tt\footnotesize \{ssh028,pneekhar\}@ucsd.edu} \\
\text{*} Equal contribution\\
}

\maketitle

\begin{abstract}
There has been a recent surge in adversarial attacks on deep learning based automatic speech recognition (ASR) systems. 
These attacks pose new challenges to deep learning security and have raised significant concerns in deploying ASR systems in safety-critical applications. 
In this work, we introduce WaveGuard: a framework for detecting adversarial inputs that are crafted to attack ASR systems. Our framework incorporates audio transformation functions and analyses the ASR transcriptions of the original and transformed audio to detect adversarial inputs.\footnote{Audio Examples: \url{https://waveguard.herokuapp.com}} We demonstrate that our defense framework is able to reliably detect adversarial examples constructed by four recent audio adversarial attacks, with a variety of audio transformation functions.
With careful regard for best practices in defense evaluations, we analyze our proposed defense and its strength to withstand adaptive and robust attacks in the audio domain. We empirically demonstrate that audio transformations that recover audio from perceptually informed representations can lead to a strong defense that is robust against an adaptive adversary even in a complete white-box setting. Furthermore, WaveGuard can be used out-of-the box and integrated directly with any ASR model to efficiently detect audio adversarial examples, without the need for model retraining.
\end{abstract}

\section{Introduction}

Speech serves as a powerful communication interface between 
humans and machine learning agents.
Speech interfaces enable hands-free operation and can assist users who are visually or physically impaired. 
Research into machine recognition of speech is driven by the prospect of offering services where humans interact naturally with machines.
To this end, automatic speech recognition (ASR) systems 
seek
to accurately convert a speech signal into 
a
transcription of the spoken words, irrespective of a speaker’s accent, or the acoustic environment in which the speaker is located~\cite{rabiner2007introduction}.
With the advent of deep learning, state-of-the-art speech recognition systems~\cite{deepspeech2,Shen2019LingvoAM,mozilladeepspeech} are based on Deep Neural Networks (DNNs) and are widely used in personal assistants 
and home electronic devices (e.g.~Apple Siri, Google Assistant).
\begin{figure}[htbp]
    \centering
    \includegraphics[width=1.0\columnwidth]{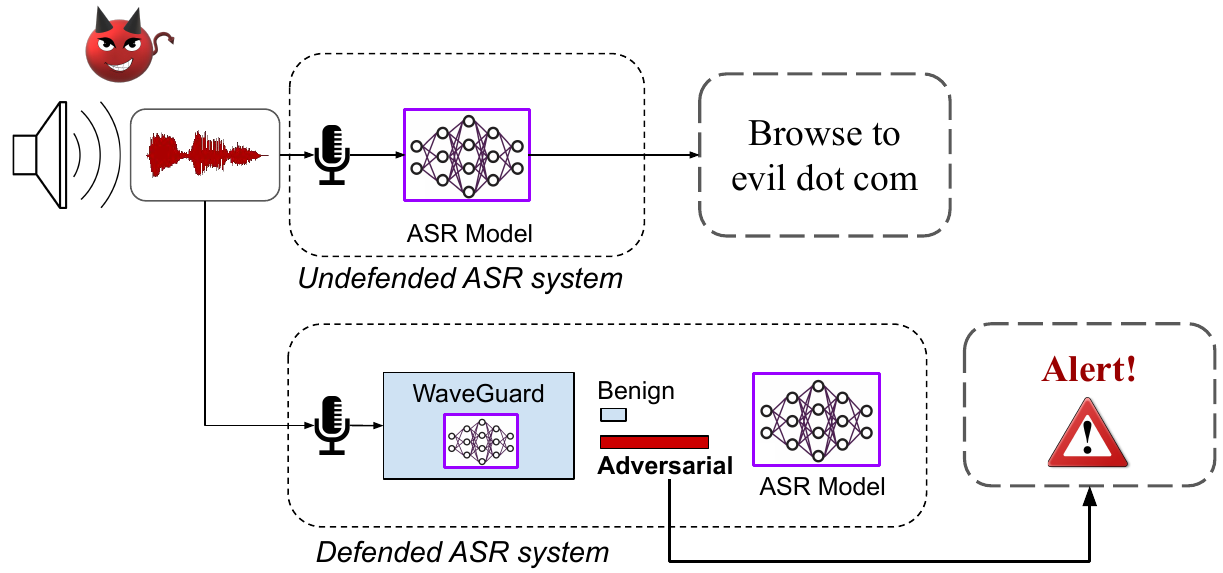}
    \caption{Depiction of an undefended ASR system and an ASR system defended by WaveGuard 
    in the presence of a malicious adversary. The ASR system defended by WaveGuard detects the adversarial input and alerts the user.}
    \label{fig:sizechart}
\end{figure}

The popularity of ASR systems has brought 
new security concerns. 
Several studies have demonstrated that DNNs are vulnerable to adversarial examples \cite{goodfellow6572explaining,obfuscated-gradients,Carlini2017TowardsET,limitations}. While previously limited to the image domain, recent attacks on ASR systems \cite{asrblack,psychoacoustic,targetattacks,hidden, DBLP:journals/corr/abs-1810-11793, qin2019imperceptible, neekhara2019universal,yuan2018commandersong,usenixdevil}, have demonstrated that adversarial examples also exist in the audio domain. An audio adversarial example can cause the original audio signal to be transcribed to a target phrase desired by the adversary or can cause significant transcription error by the victim ASR model. 

Due to the existence of these vulnerabilities, there is a crucial need for
defensive methods that can be employed to thwart audio adversarial attacks.
In the image domain, several works have proposed input transformation based defenses~\cite{meng2017magnet,guo2018countering,lindefensive,khalid2019qusecnets,liang2018detecting} to recover benign images from adversarially modified images. Such inference-time adversarial defenses use image transformations like feature squeezing, JPEG compression, quantization, randomized smoothing (etc.)~to render adversarial examples ineffective. While such defenses are effective in guarding against non-adaptive adversaries, they can be bypassed in an adaptive attack scenario where the attacker has partial or complete knowledge about the defense. 

Another line of defense in the image domain is based on training more robust neural networks using adversarial training or by introducing randomization in network layers and parameters. 
Such defenses are comparatively more robust under adaptive attack scenarios, however they are significantly more expensive to train as compared to input transformation based defenses that can be employed directly at the model inference stage. 
Although input transformation based defenses are shown to be broken for image classifiers, the same conclusion cannot be drawn for ASR systems without careful evaluation. This is because an ASR system is a more complicated architecture as compared to an image classification model and involves several individual components: an acoustic feature extraction pipeline, a neural sequence model for processing the time-series data and a language head for predicting the language tokens. This pipeline makes it challenging to craft robust adversarial examples for ASR systems that can reliably transcribe to a target phrase even when the input is transformed and reconstructed from some perceptually informed representation.



\textbf{WaveGuard:} In this work, we study the effectiveness of audio transformation based defenses for detecting adversarial examples for speech recognition systems. We first design a general framework for employing audio transformation functions as an adversarial defense for ASR systems.
Our framework transforms the given audio input $x$ using an input transformation function $g$ and analyzes the ASR transcriptions for the input $x$ and $g(x)$. The underlying idea for our defense is that model predictions for adversarial examples are unstable while those for benign examples are robust to small changes in the input. Therefore, our framework labels an input as adversarial if there is a significant difference between the transcriptions of $x$ and $g(x)$.

We first study five different audio transformations under different compression levels against non-adaptive adversaries. We find that at optimal compression levels, most input transformations can reliably discriminate between adversarial and benign examples for both targeted and untargeted adversarial attacks on ASR systems. Furthermore, we achieve higher detection accuracy in comparison to prior work~\cite{ensembleaudiodefense,yang2019characterizing} in adversarial audio detection. However, this evaluation does not provide security guarantees against a future adaptive adversary who has knowledge of our defense framework. To evaluate the robustness of our defense against an adaptive adversary, we propose a strong white-box adaptive attack against our proposed defense framework. Interestingly, we find that some input transformation functions are robust to adaptive attack even when the attacker has complete knowledge of the defense. Particularly, the transformations that recover audio from perceptually informed representations of speech prove to be more effective against adaptive-attacks than naive audio compression and filtering techniques.\\

\noindent \textbf{Summary of Contributions:} 
\begin{itemize}
    
    \item We develop a formal defense framework (Section~\ref{sec:methodology}) for detecting audio adversarial examples against ASR systems. Our framework uses input transformation functions and analyses the transcriptions of original and transformed audio to label the input as adversarial or benign.
    \item We evaluate different transformation functions for detecting recently proposed and highly successful targeted~\cite{targetattacks,qin2019imperceptible} and untargeted~\cite{neekhara2019universal} attacks on ASR systems.
    We study the trade-off between the hyper-parameters of different transformations and the detector performance and find an optimal range of hyper-parameters for which the given transformation can reliably detect adversarial examples (Section~\ref{sec:nonadaptiveevals}). 
    
    
    \item We demonstrate the robustness of our defense framework against an adaptive adversary who has complete knowledge of our defense and intends to bypass it. We find that certain input transformation functions that reduce audio to a perceptually informed representation cannot be easily bypassed under different allowed magnitudes of perturbations. Particularly, we find that Linear Predictive Coding (LPC) and Mel spectrogram extraction-inversion are more robust to adaptive attacks as compared to other transformation functions studied in our work (Section~\ref{sec:adaptivemethodology}). 
    
    \item We investigate transformation functions for the goal of recovering the original transcriptions from an adversarial signal. We find that for certain attacks and transformation functions, we can recover the original transcript with a low Character Error Rate. (Section~\ref{sec:transcriptevals})
\end{itemize}

\section{Background and Related Work}
\subsection{Adversarial Attacks in the Audio Domain: }

Adversarial attacks on ASR systems have primarily focused on \textit{targeted attacks} to embed carefully crafted perturbations into speech signals, such that the victim model transcribes the input audio into a specific malicious phrase, as desired by the adversary \cite{asrblack,targetattacks,mfccattack, hidden,usenixaudio}. Such attacks can for example cause a digital assistant to incorrectly recognize commands it is given, thereby compromising the security of the device. Prior works~\cite{hidden,usenixaudio} demonstrate successful attack algorithms targeting traditional speech recognition models based on HMMs and GMMs~\cite{baum1967,baum1970maximization,hmm1,ahadi1997combined,bahl1986maximum,Rabiner89-ATO}. For example, in Hidden Voice Commands~\cite{hidden}, the attacker uses inverse feature extraction to generate obfuscated audio that can be played over-the-air to attack ASR systems. However, obfuscated samples sound like random noise rather than normal human perceptible speech and therefore come at the cost of being fairly perceptible to human listeners. 

In more recent work~\cite{targetattacks} involving neural network based ASR systems, Carlini \emph{et al.}~propose an end-to-end white-box attack technique to craft adversarial examples, which transcribe to a target phrase. Similar to work in images, they propose a gradient-based optimization method that replaces the cross-entropy loss function used for classification, with a Connectionist Temporal Classification (CTC) loss~\cite{graves2006connectionist} which is optimized for time-sequences. The CTC-loss between the target phrase and the network's output is backpropagated through the victim neural network and the Mel Frequency Cepstral Coefficient (MFCC) computation, to update the additive adversarial perturbation. The authors in this work demonstrate 100\% attack success rate on the Mozilla DeepSpeech~\cite{mozilladeepspeech} ASR model. The adversarial samples generated by this work are quasi-perceptible, motivating a separate work~\cite{psychoacoustic} to minimize the perceptibility of the adversarial perturbations using psychoacoustic hiding. Further addressing the imperceptibility of audio attacks, Qin \emph{et al.}~\cite{qin2019imperceptible} develop effectively imperceptible audio adversarial examples by leveraging the psychoacoustic principle of auditory masking. In their work~\cite{qin2019imperceptible}, the imperceptibility of adversarial audio is verified through a human study, while retaining 100\% targeted attack success rate on the Google Lingvo~\cite{Shen2019LingvoAM} ASR model.

Targeted attacks, such as those described above, cannot be performed in real-time since it requires the adversary to solve a data-dependent optimization problem for each data-point they wish to mis-transcribe. To perform attacks in real-time, the authors of~\cite{neekhara2019universal} designed an algorithm to find a single quasi-imperceptible universal perturbation, which when added to any arbitrary speech signal, causes mis-transcription by the victim speech recognition model. The proposed algorithm 
iterates
over the training dataset to build a universal perturbation vector, that can be added to any speech waveform to cause an error in transcription by a speech recognition model with high probability. This work also demonstrates transferability of adversarial audio samples across two different ASR systems (based on DeepSpeech and Wavenet), demonstrating that such audio attacks can be performed in real-time even when the attacker does not have knowledge of the ASR model parameters.

\begin{figure}[htbp]
    \centering
    \includegraphics[width=1.0\columnwidth]{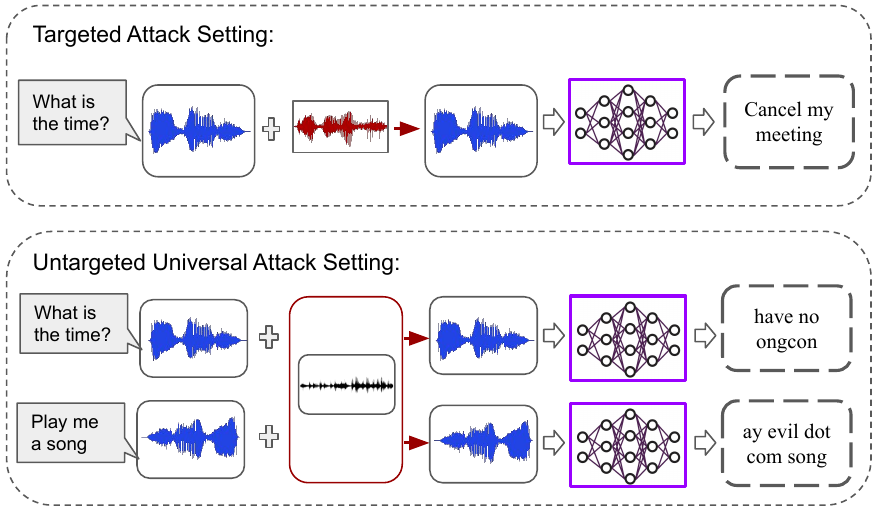}
    
    \caption{\textbf{Top:} In the targeted attack setting, the adversary solves a data-dependent optimization problem to find an additive perturbation, such that a victim ASR model transcribes the adversarial input audio to a target phrase as desired by the adversary. 
    \textbf{Bottom:} In the untargeted universal attack setting, the adversary computes a single universal perturbation which when added to any arbitrary audio signal, will most likely cause an error in transcription by a victim ASR system. In untargeted attacks, the transcription of adversarial audio may not be a specific malicious phrase.}
    \label{fig:generalpipeline}
\end{figure}

\noindent \textbf{Physical attacks.} Adversarial attacks to ASR Systems have also been demonstrated to be a real-world threat. In particular, recently developed attack algorithms have shown success in attacking physical  intelligent voice control (IVC) devices, when playing the generated adversarial examples over-the-air. The recently developed \emph{Devil's Whisper}~\cite{usenixdevil} demonstrated that adversarial commands embedded in music samples and played over-the-air using speakers, are able to attack popular IVC devices such as Google Home, Google Assistant, Microsoft Cortana and Amazon Alexa with 98\% of target commands being successful. They utilize a surrogate model approach to generate transferable adversarial examples that can attack a number of unseen target devices. 
However, as noted by the authors, physical attacks are very sensitive to various environmental factors, such as the volume when playing adversarial examples, the distance between the
speaker and the victim IVC device, as well as the brand of speakers, that can render the attack unsuccessful. 
Qin \emph{et al.}~\cite{qin2019imperceptible} designed robust, physical-world, over-the-air audio adversarial examples by constructing perturbations, which remain effective in attacking the Google Lingvo ASR model~\cite{Shen2019LingvoAM} even after applying
environmental distortions. Such robust adversarial examples are crafted by incorporating the noise simulation during the training process of the perturbation. In our work, we evaluate our defense against the robust attack proposed in~\cite{qin2019imperceptible} on the Google Lingvo ASR model. 
We find that while such examples are more robust to small input changes as compared to previously proposed targeted attacks~\cite{targetattacks}, 
they can still be easily distinguished from benign audio samples using our defense framework. 


\subsection{Principles of Defense and Adaptive Attacks in the Image Domain}
\label{sec:related_adaptive}
To strengthen the reliability of deep learning models in the image domain, a significant amount of prior work has proposed defenses to adversarial attacks~\cite{meng2017magnet,guo2018countering,liang2018detecting,lindefensive,Qin2020Detecting,xie2018mitigating}.
However, most of these defenses were only evaluated against non-adaptive attacks or using a ``zero-knowledge'' threat model, where the attacker has no knowledge of the defense existing in the system.
Such defenses offer bare-minimum security and in no way guarantee that they can be secure against future attacks~\cite{biggio2013evasion,carlini2019evaluating}.
Accurately evaluating the robustness of defenses is a challenging but important task, particularly because of the presence of adaptive adversaries~\cite{obfuscated-gradients,carlini2019evaluating,carlini2017adversarial,tramer2020adaptive}. An adaptive adversary is one that has partial or complete knowledge of the defense mechanism in place and therefore adapts their attack to what the defender has designed~\cite{carlini2019evaluating,herley2017sok,carlini2017adversarial}. 

Many
prior works on defenses are variants of the same idea: pre-process inputs using a transform, e.g.~randomized cropping, rotation, JPEG compression, randomized smoothing, auto-encoder transformation, that can remove the adversarial perturbation from the input. However, such defenses are shown to be vulnerable to attack algorithms that are partially or completely aware of the defense mechanism~\cite{obfuscated-gradients,eot}. In~\cite{obfuscated-gradients}, the authors show that the input-transformation function can be substituted with a differentiable approximation in the backward pass in-order to craft adversarial examples that are robust under the given input-transform. In~\cite{eot}, the authors craft adversarial examples that are robust over a given distribution of transformation functions, which guarantees robustness over more than one type of transform. 

Solely analyzing a defense against a non-adaptive adversary gives us a false sense of security. Therefore, the authors of~\cite{carlini2019evaluating} provided several guidelines to ensure completeness in the evaluation of defenses to adversarial attacks. The authors recommend using a threat model with an ``infinitely thorough'' adaptive adversary, who is capable of developing new optimal attacks against the proposed defense. They recommend applying a diverse set of attacks to any proposed defense, with the same mindset of a future adversary. However, such defense guidelines have not been applied to the audio domain and many of the proposed ASR defenses have not carried out thorough evaluations against adaptive adversaries. In our work, we follow these guidelines and evaluate our ASR defense against the strongest non-adaptive and adaptive adversaries.

\subsection{Defenses in the Audio Domain}
In comparison to the image domain, only a handful of studies have proposed defenses to adversarial attacks in the audio domain.
Prior work on defenses for speech recognition models have focused on both audio pre-processing techniques~\cite{ensembleaudiodefense,detectingposter} and utilizing temporal dependency in speech signals~\cite{yang2019characterizing} to detect adversarial examples.

Yang \emph{et al.}
in~\cite{yang2019characterizing} proposed a defense framework against three attack methods targeting state-of-the-art ASR models such as Kaldi and DeepSpeech. The proposed defense framework checks if the transcription of the first $k$-sized portion of the audio waveform ($t_{\mathit{1}}$) is similar to the first $k$-sized transcription of the complete audio waveform ($t_{\mathit{2}}$). 
A sample is identified as adversarial when the two transcriptions are dissimilar, i.e.,~the Character Error Rate (CER) or Word Error Rate (WER) between $t_{\mathit{1}}$ and $t_{\mathit{2}}$ is higher than a predefined threshold. 
The authors further study the effectiveness of their defense in an adaptive attack scenario, where the attacker has partial knowledge of the defense framework. 
In their strongest adaptive attack scenario, they vary the portion $k_{\mathit{D}}$ used by the defense and evaluate the cases where the adaptive attacker uses a
the same/different portion $k_{\mathit{A}}$. 

However, recent work~\cite{tramer2020adaptive} has re-evaluated temporal dependency frameworks and demonstrated them to be ineffective in detecting adversarial perturbations in the audio domain. The authors of~\cite{tramer2020adaptive} designed attacks that were able to fool the proposed detector in~\cite{yang2019characterizing} with 100\% accuracy, and further report that the adaptive evaluations conducted in~\cite{yang2019characterizing} are incomplete. In the adaptive attack designed by~\cite{tramer2020adaptive}, the CTC loss function used by the attacker incorporates different values of $k_{\mathit{A}}$ and is therefore able to bypass the temporal dependency detector with minimal added perturbation to audio.

Aside from proposing the temporal-dependency defense for detection, the authors of~\cite{yang2019characterizing} also study the effectiveness of various input transformation functions in recovering the original transcription from the adversarial counterpart. To this end, they perform experiments with transformation functions such as quantization, down-sampling, local smoothing and auto-encoder reformation of signals. They report that these methods are ineffective in recovering the correct transcription of audio signals. In our work, we will 
evaluate
some of these transformations for the goal of detecting adversarial examples as opposed to recovering benign examples. However, we report that for some attack types, most transformation based defenses are able to recover the benign audio transcription with low CER.  

Rajaratnam \emph{et al.}~\cite{ensembleaudiodefense} also studied the use of pre-processing techniques such as audio compression, band-pass filtering, audio panning and speech coding as a part of both isolated and ensemble methods for detecting adversarial audio examples generated by a single targeted attack~\cite{carlini2017adversarial}. While they report high detection performance against the targeted adversarial attack proposed by~\cite{carlini2017adversarial}, their techniques were not evaluated in an adaptive attack setting and therefore do not provide security guarantees against a future adversary. 
Given the difficulty of performing defense evaluations, in our work, we perform additional experiments with various input transformation functions to validate or refute the security claims made in existing papers. 


\begin{figure*}[htbp]
    \centering
    \includegraphics[width=0.87\textwidth]{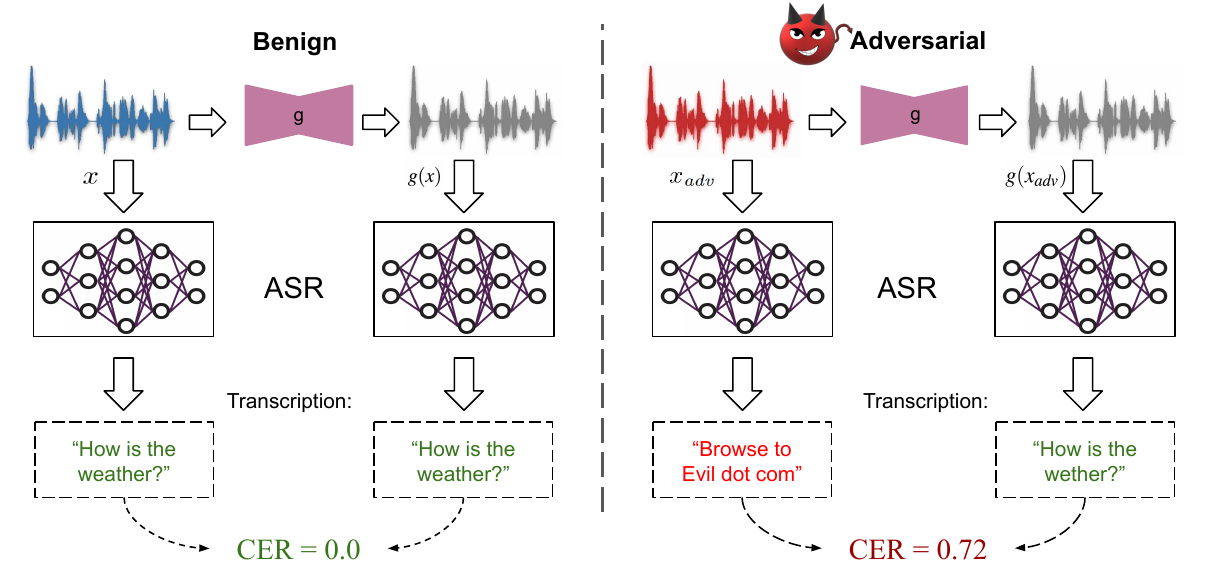}
    \caption{WaveGuard Defense Framework: We first processes the input audio $x$ using an audio transformation function $g$ to obtain $g(x)$. Then the ASR transcriptions or $x$ and $g(x)$ are compared. An input is classified as \textit{adversarial} if the difference between the transcriptions of $x$ and $g(x)$ exceeds a particular threshold. }
    \vspace{-4mm}
    \label{fig:overview}
\end{figure*}

\section{Methodology}
\label{sec:methodology}
\subsection{Threat Model}
Adversarial attacks in the audio domain can be classified broadly into two categories: \textit{targeted} and \textit{untargeted} attacks. In targeted attacks the goal of the adversary is to add a small perturbation to an audio signal such that it causes the victim ASR to transcribe the audio to a given target phrase. In untargeted attacks the goal is simply to cause significant error in transcription of the audio signal so that the original transcription cannot be deciphered. 

The common goal across both targeted and untargeted attack is to cause mis-transcription of the given speech signal while keeping the perturbation 
imperceptible. 
Therefore, we define an audio adversarial example $x_{\mathit{adv}}$ as a perturbation of an original speech signal $x$ such that the Character Error Rate (\textit{CER}) between the transcriptions of the original and adversarial examples from an ASR $C$ is greater than some threshold $t$. That is,
\begin{equation}
\mathit{CER}( C(x), C(x_{\mathit{adv}}) ) > t
\end{equation}
and the distortion between $x_{\mathit{adv}}$ and $x$ is constrained under a distortion metric $\delta$ as follows:
\begin{equation}
\delta(x, x_{\mathit{adv}}) < \epsilon.
\end{equation}
Here, $\mathit{CER}(x, y)$ is the edit distance~\cite{editdistance} between the strings $x$ and $y$ normalized by the length of the strings i.e.,
\begin{equation}
\mathit{CER}(x,y) = \frac{\mathit{EditDistance}(x, y)}{\textit{max}(\mathit{length}(x), \mathit{length}(y))}.
\end{equation}

$L_p$ norms are popularly used to quantify the distortion $\delta$ between the original and adversarial example in the image domain. Following prior works~\cite{targetattacks,neekhara2019universal} on audio adversarial attacks, we use an $L_\infty$ norm on the waveforms to quantify the distortion between the adversarial and the original signal. 


\subsection{Defense Framework}
\label{sec:defenseframework}
The goal of our defense is to correctly detect adversarially modified inputs. 
The underlying hypothesis for our defense framework is that the network predictions for adversarial examples are often unstable and small changes in adversarial inputs can cause significant changes in network predictions. In the image domain, it has been shown that several input transformation techniques~\cite{meng2017magnet,guo2018countering,lindefensive,khalid2019qusecnets} such as JPEG compression, randomized smoothing and feature squeezing can
render
adversarial perturbations ineffective. This is because such input transformations introduce an additional perturbation in the input that can dominate the carefully added adversarial perturbation.
On the other hand, predictions for the original (benign) inputs are usually robust to small random perturbations in the input.

Based on this hypothesis, we propose the following defense framework for detecting audio adversarial examples: For a given audio transformation function $g$, input audio $x$ is classified as adversarial if there is significant difference between the transcriptions $C(x)$ and $C(g(x))$:
\begin{equation}
\mathit{d}(C(x), C(g(x))) > t
\end{equation}
where $d$ is some distance metric between the two given texts and $t$ is a detection threshold. In our work we use the Character Error Rate (CER) as the distance metric $d$. z
An overview of the defense is depicted in Figure~\ref{fig:overview}. Note that unlike~\cite{yang2019characterizing}, the goal using an input transformation $g$ is not to recover the original transcription of an adversarial example, but to detect if an example is adversarial or benign by observing the difference in the transcriptions of 
$x$ and $g(x)$.

In this work, we study various input transformation functions $g$ as candidates for our defense framework. We evaluate our defense against four recent adversarial attacks~\cite{carlini2017adversarial,qin2019imperceptible,neekhara2019universal} on ASR systems. 
One of the main insights we draw from our experiments is that in the non-adaptive attack setting, most audio transformations can be effectively used in our defense framework to accurately distinguish adversarial and benign inputs. This result is consistent with the success of input-transformation based defenses in the image domain.

\begin{figure*}[h]
    \includegraphics[width=1.0\textwidth]{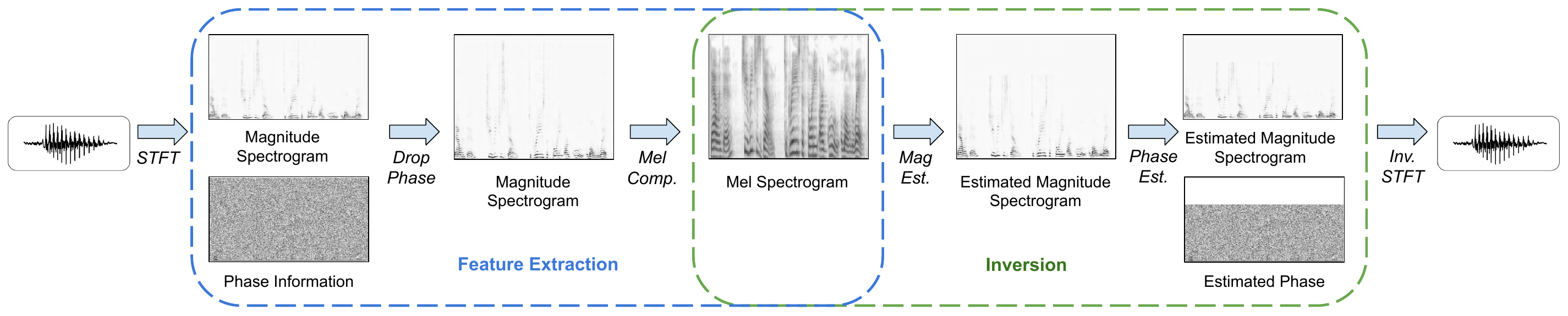}
    
    \caption{Steps involved in the Mel extraction and inversion transform (Section \ref{sec:melinversion}). In the extraction step, the phase information of the signal is discarded and the magnitude spectrogram is compressed to a Mel spectrogram using a linear transform. In the inversion step, the waveform is estimated by first estimating the magnitude spectrogram, followed by phase estimation and finally an inverse STFT.}
    \label{fig:extract_invert}
\end{figure*}

However, in order to use a defense reliably in practice, the defense must be secure against an adaptive adversary who has knowledge of the defense. 
For an adaptive attack setting, we find that certain input transformations are more robust to attacks than others. Particularly, the transformations which compress audio to perceptually informed representations cannot be easily bypassed even when the attacker has complete knowledge of the defense. 
This finding is in contrast to the image domain where most input transformation based defenses have been shown to be broken under robust or adaptive adversarial attacks. 
We elaborate on our adaptive attack scenario and the results in Section~\ref{sec:adaptivemethodology} and Section~\ref{sec:adaptiveresults}.


\section{Input-transformation functions}
\label{sec:transformations}

We study the following audio transformations as candidates for the input transformation function $g$:

\subsection{Quantization-Dequantization}
Several works in the image domain~\cite{khalid2019qusecnets,Lu_2017_ICCV,xu2017feature}, have used quantization based defenses to neutralize the effect of adversarial perturbations. Since adversarial pertubations to audio have small amplitudes, quantization can help reomve added perturbations. In this study, we employ  quantization-dequantization in our defense framework, where each waveform sample is quantized to $q$ bits and then reconstructed back to floating point to produce the output approximation of the original input data.

\subsection{Down-sampling and Up-sampling} 
Discarding samples from a waveform during down-sampling could remove a significant portion of the adversarial perturbation, thereby disrupting an attack. To study this effect, we down-sample the original waveform (16 kHz in our experiments), to a lower sampling rate and then estimate the waveform at its original sampling rate using interpolation. We perform this study for a number of different down-sampling rates to find an optimal range of sampling rates for which the defense is effective.

\subsection{Filtering}
\label{sec:filtering}
Filtering is commonly applied for noise cancellation applications such as removing background noise from a speech signal. 
It is intuitive to study the effect of filtering in order to remove adversarial noise from a speech signal. 
In this work, we use low-shelf and high-shelf filters to clean a given signal. Low-shelf and high-shelf filters are softer versions of high-pass and low-pass filters respectively. That is, instead of completely removing frequencies above or below some thresholds, shelf filters boost or reduce their amplitude. For noise removal, we use a low-shelf filter to reduce the amplitude of frequencies below a threshold and a high-shelf filter to reduce the amplitude of frequencies above a threshold. 

In our experiments we first compute the spectral centroid of the audio waveform: Each frame of a magnitude spectrogram is normalized and treated as a distribution over frequency bins, from which the mean (centroid) is extracted per frame. We then compute the median centroid frequency ($C$) over all frames and set the high-shelf frequency threshold as $1.5\times C$ and low-shelf frequency threshold as $0.1 \times C$. We then reduce the amplitude of frequencies above and below the respective thresholds using a negative gain parameter of -30.

\subsection{Mel Spectrogram Extraction and Inversion}
\label{sec:melinversion}
Mel spectrograms are popularly used as an intermediate audio representation in both text-to-speech~\cite{shen2018natural,neekhara2019expediting,miao2020flow} and speech-to-text~\cite{mfccASR,kaldipytorch} systems. While reduction of the waveform to a Mel spectrogram is a lossy compression, the Mel spectrogram is a perceptually informed representation that mostly preserves the audio content necessary for speech recognition systems. We use the following Mel spectrogram extraction and inversion pipeline for disrupting adversarial perturbations in our experiments: 

\textbf{Extraction:}
We first decompose waveforms into time and frequency components using a Short-Time Fourier Transform (STFT). Then,
the phase information is discarded from the complex STFT coefficients leaving only the magnitude spectrogram. 
The linearly-spaced frequency bins of the resultant spectrogram are then compressed to fewer bins which are equally-spaced on a logarithmic scale (usually the Mel scale~\cite{stevens1937scale}). 
Finally, amplitudes of the resultant spectrogram are made logarithmic to conform to human loudness perception, then optionally clipped and normalized to obtain the Mel spectrogram.

\textbf{Inversion:}
To invert the Mel spectrogram into a listenable waveform, 
the inverse of each extraction step is applied in reverse.
First, logarithmic amplitudes are converted to linear ones. 
Then the magnitude spectrogram is estimated from the Mel spectrogram using the approximate inverse of the Mel transformation matrix. 
Next, the phase information is estimated from the magnitude spectrogram using a heurisitc algorithm such as Local Weighted Sum (LWS)~\cite{lws} or Griffin Lim~\cite{griffinlim}. Finally, the inverse STFT is used to render audio from the estimated magnitude spectrogram and phase information.

We hypothesize that reconstructing audio from a perceptually informed representation can potentially remove the adversarial perturbation while preserving the speech content that is perceived by the human ear. While some speech recognition systems also use Mel spectrogram features, we find that reconstructing audio from the \textit{compressed} Mel spectrograms introduces enough distortion in the original waveform, such that the ASR Mel features of the newly reconstructed audio are different from the original audio. The distortion in the reconstructed audio is introduced by the magnitude estimation and phase estimation steps depicted in Figure~\ref{fig:extract_invert}.
In order to bypass a defense involving Mel extraction and inversion, an adaptive attacker will need to craft a perturbation that can be retained in the compressed Mel spectrogram representation, making it challenging to keep the perturbation imperceptible. In our adaptive attack experiments in Section~\ref{sec:adaptiveresults} we demonstrate that even when the attacker uses a differentiable implementation of the Mel extraction and inversion pipeline, it cannot easily be bypassed without introducing a 
clearly
perceptible adversarial noise in the signal.



\subsection{Linear Predictive Coding}
\label{sec:lpc}
\begin{figure}[htp]
    \centering
    \includegraphics[width=1.0\columnwidth]{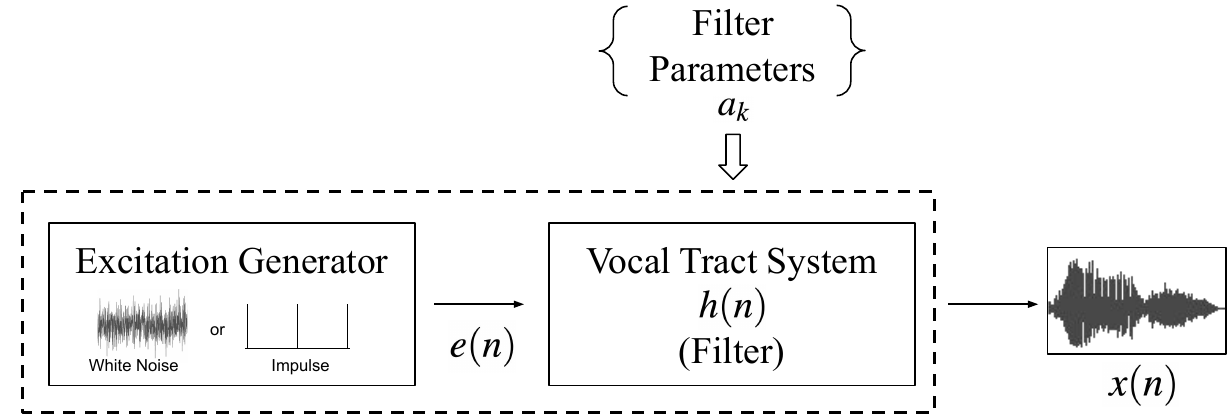}
    
    \caption{Model for linear predictive analysis of speech signals.}
    \label{fig:lpcfig}
\end{figure}

Linear Predictive Coding (LPC) is a speech encoding technique that uses a source-filter model based on a
mathematical approximation of the human vocal tract. The model assumes that a source signal $e(n)$ (which models the vocal chords) is passed as input to a resonant filter $h(n)$ (that models the vocal tract) to produce the resultant signal $x(n)$.  That is:
\begin{equation}
x(n) = h(n) * e(n)
\end{equation}
The source excitation $e(n)$ can either be quasi-periodic impulses (during voiced speech) or random noise (during unvoiced speech). Both these source excitation sources are spectrally flat implying that all spectral information is modeled in the filter parameters.

LPC assumes a 
$p^{\text{th}}$ order all-pole filter $h(n)$ which means that each waveform sample is modelled as a linear combination of $p$ previous values. That is, 
\begin{equation}
x(n) = \Sigma_{k=1}^{k=p}a_k x(n-k) + e(n). 
\end{equation}
The basic problem of LPC analysis is to estimate the filter parameters $a_k$. Since the source signal is assumed to be an impulse train or random white noise, the problem is formulated as minimizing $||e(n)||^2$ which is the power of the excitation signal. This reduces the parameter-estimation problem to a linear regression problem 
in which the goal is to minimize:
\begin{equation}
\textit{minimize: } \langle ||e(n)||^2 \rangle = \langle(x(n) -  \Sigma_{k=1}^{k=p}a_k x(n-k))^2 \rangle 
\label{eq:lpclinearregression}
\end{equation}

Here, $\langle \rangle$ denotes averaging over finite number of waveform samples.
In practice, a long time-varying signal is divided into overlapping windows of size $w$ and LPC coeffecients $a_k$ are estimated for each window by solving the above 
linear regression problem. To re-synthesize the signal from the estimated coefficients, we use a random-noise excitation signal. In our experiments, we use 25 millisecond windows with 12.5 millisecond overlap. We experiment with different numbers of the LPC coeffecients which control the compression level of the original signal. 

Since LPC models the human vocal tract system, it preserves the phonetic information of speech in the filter parameters. Bypassing a defense involving LPC transform, would require the adversary to add an adversarial perturbation that can be preserved in the LPC filter coeffecients; thereby requiring the adversary to modify the phonetic information in speech. We empirically demonstrate that the LPC transform cannot be easily bypassed by an adaptive adversary. 

\section{Experimental Setup}

We evaluate our defense against the following recent audio adversarial attacks on speech recognition systems~\cite{targetattacks,neekhara2019universal,qin2019imperceptible}:
\begin{itemize}
    \item \textbf{Carlini:} Attack introduced in~\cite{targetattacks}.
    This is a white-box targeted attack on the Mozilla Deepspeech~\cite{mozilladeepspeech} ASR system, where the attacker trains an adversarial perturbation by minimizing the CTC loss between the target transcription and the ASR's prediction. This attack minimizes the $L_\infty$ norm of the adversarial perturbation to constrain the amount of distortion.
    
    \item \textbf{Qin-I:}  Imperceptible attack described in~\cite{qin2019imperceptible}. This is another white-box targeted attack that focuses on ensuring imperceptibility of the adversarial perturbation by using psycho-acoustic hiding. The victim ASR for this attack is Google Lingvo~\cite{Shen2019LingvoAM}.
    
    \item \textbf{Qin-R:} Robust attack described in~\cite{qin2019imperceptible}. This attack incorporates input transformations during training of the adversarial perturbation which simulate room environments. This improves the attack robustness in real world settings when played over the air. The victim ASR for this attack is Google Lingvo~\cite{Shen2019LingvoAM}.
    
    \item \textbf{Universal:} We implement the white-box attack described in~\cite{neekhara2019universal}. This is an untargeted attack which finds an input-agnostic perturbation that can cause significant disruption in the transcription of the adversarial signal. In our work, we follow the algorithm provided by the authors and craft universal perturbation with an $L_\infty$ bound of 400 (for 16-bit audio wave-forms with sample values in the range -32768 to 32768). The victim ASR for this attack is Mozilla DeepSpeech~\cite{mozilladeepspeech}.
\end{itemize}

\begin{table}[htbp]
\centering
\resizebox{0.75\columnwidth}{!}{%
\begin{tabular}{@{}cl@{}}
\toprule
\multicolumn{2}{c}{\textbf{Target Adversarial Commands}} \\ \midrule
\multicolumn{2}{c}{"browse to evil dot com"} \\
\multicolumn{2}{c}{"hey google cancel my medical appointment"} \\
\multicolumn{2}{c}{"hey google"} \\
\multicolumn{2}{c}{"this is an adversarial example"} \\ \bottomrule
\end{tabular}%
}
\caption{Adversarial commands used for constructing targeted adversarial examples.}
\label{tab:targetphrase}
\end{table}

\begin{table*}[h]
\centering
\resizebox{1.0\textwidth}{!}{%
\begin{tabular}{l|c|cccc|cccc}
\multicolumn{1}{c}{} &
\multicolumn{1}{c}{} &
\multicolumn{4}{c}{\emph{AUC Score}} &
\multicolumn{4}{c}{\emph{Detection Accuracy}}\\
\toprule
Defense & Hyper-params & Carlini & Universal & Qin-I & Qin-R & Carlini & Universal & Qin-I & Qin-R\\
\midrule
Downsampling - Upsampling & 6000 kHz & 1.00 & 0.91 & 1.00 & 1.00 & 100\%  & 88\% & 100\% & 100\%\\
Quantization - Dequantization & 6 bits & 0.99 & 0.92 & 1.00 & 0.93 & 98.5\% & 88\% & 99\% & 95\%\\
Filtering  & (Section~\ref{sec:filtering}) & 1.00 & 0.92 & 1.00 & 1.00 & 99.5\% & 86\% & 100\% & 100\%\\
Mel Extraction - Inversion & 80 Mel-bins & 1.00 & 0.97 & 1.00 & 1.00 & 100\% & 92\% & 100\% & 100\%\\
LPC & LPC order 20 & 1.00 & 0.91 & 1.00 & 1.00 & 100\% & 83\% & 100\% & 100\% \\
\bottomrule
\end{tabular}
}
\caption{Evaluations for each input transformation defense against various non-adaptive attacks. We use two objective metrics: AUC score and Attack Detection Accuracy for evaluation (higher values are better for both metrics). }
\label{tab:nonadaptiveresult}
\end{table*}

\subsection{Dataset and Attack Evaluations}
We conduct all our experiments on the Mozilla Common Voice dataset, which contains 582 hours of audio across 400,000 recordings in English. The audio data is sampled at 16 kHz. We evaluate on the same subset of the Mozilla Common Voice dataset, as used in~\cite{targetattacks}, that is, the first 100 examples from the Mozilla Common Voice test set. We construct adversarial examples on this dataset using each of the attacks described above. In the targeted attack scenario, we randomly choose one of the target phrases listed in Table~\ref{tab:targetphrase} and follow the attack algorithms to create 100 pairs of original and adversarial examples for each attack type. For the untargeted universal attack, we train the universal perturbation on the same subset of Mozilla Common Voice examples with $L_\infty$ distortion bound of 400.

\noindent \textbf{Attack evaluations:} We achieve 100\% attack success rate for \textit{Carlini} and \textit{Qin-I} attacks. For \textit{Qin-R}, the attack achieves 47\% success rate (similar to that reported in the paper~\cite{qin2019imperceptible}) on 100 examples. In our experiments when recreating the \textit{Universal} attack, we achieve an attack success rate of 81\% using the same criteria as described in~\cite{neekhara2019universal} i.e., the attack is considered successful when the CER between original and adversarial transcriptions is greater than 0.5.  

\subsection{Evaluation Metrics}
As described in Section~\ref{sec:defenseframework}, in our detection framework, we label an example as \textit{adversarial} or \textit{benign} based on the CER between $x$ and $g(x)$. The decision threshold $t$ controls the true positive rate and false positive rate of our detector. Following standard procedure to evaluate such detectors~\cite{yang2019characterizing}, we calculate the \textit{AUC score} - Area Under the ROC curve. A higher AUC score indicates that the detector has more discriminative power against adversarial examples.

Additionally, we also report the \textit{Detection Accuracy} which is calculated by finding the best detection threshold $t$ on a separate set containing 50 adversarial and benign examples.

\begin{figure}[htp]
    \centering
    \includegraphics[width=1.0\columnwidth]{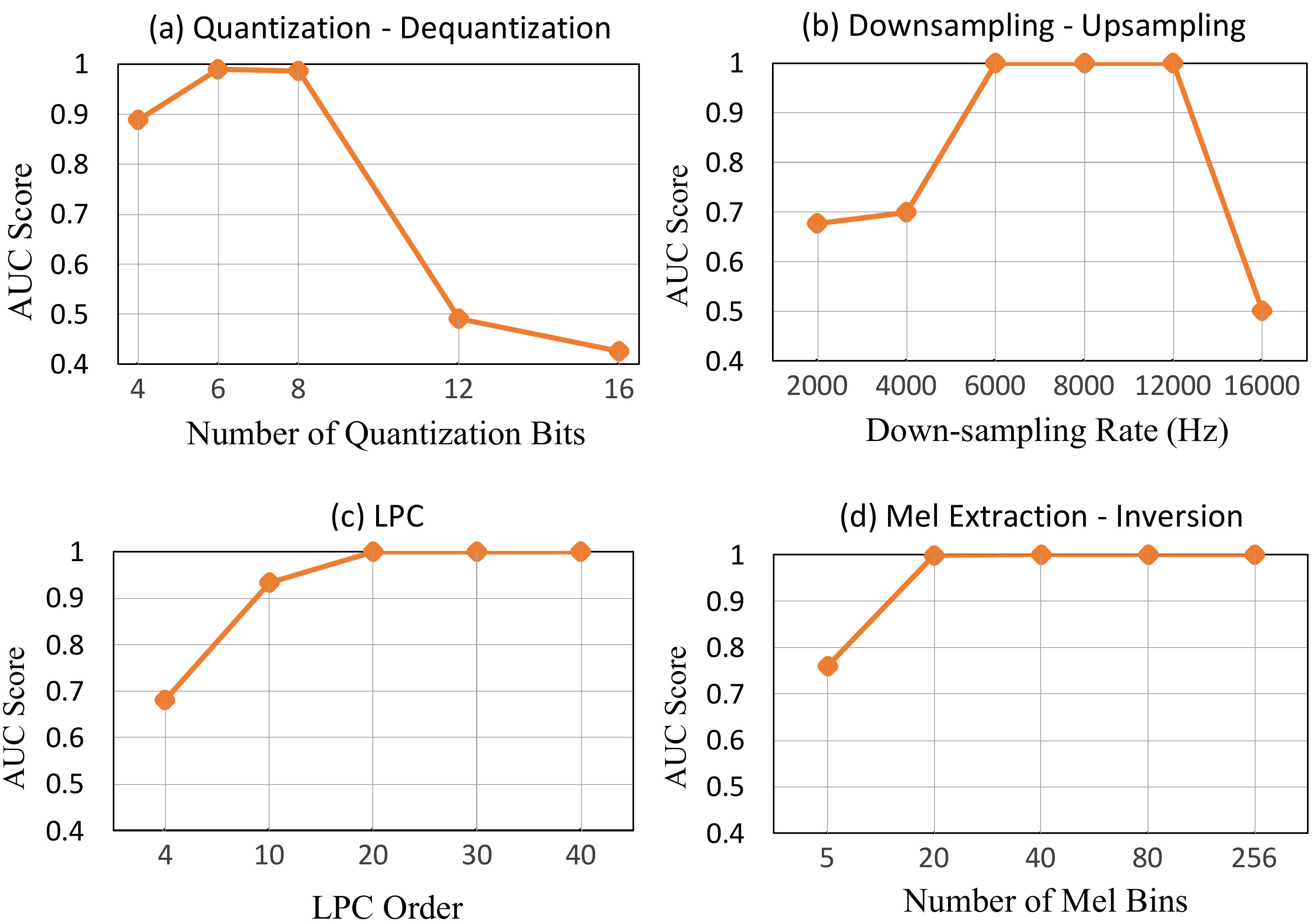}
    
    \caption{Detection AUC Scores against \textit{Carlini} attack at varying compression levels for the following transforms: (a) Quantization - Dequantization; (b) Downsampling - Upsampling; (c) Linear Predictive Coding (LPC); and (d) Mel Spectrogram Extraction- Inversion. }
    \label{fig:nonadaptive_plots}
\end{figure}

\section{Evaluation against Non Adaptive Attacks}
\label{sec:nonadaptiveevals}

The various input transformation functions we consider can be parameterized to control the compression level of the transformation.
There is a trade-off between the compression level and the discriminative power of the detector.
At low compression levels the transformation may not 
eliminate
the adversarial perturbation. In contrast, at very high compression levels, even the benign signals may become significantly distorted causing substantial change in their transcriptions. Keeping this in mind, we perform a search over the hyper-parameters for different audio transforms.
The AUC score of the detector against the \textit{Carlini} attack for different transformation functions at varying compression levels is depicted in Figure \ref{fig:nonadaptive_plots}. For most transformations, we observe the expected pattern where the defense is effective at some optimal compression levels and the AUC falls at very high or low compression levels. 
The Mel extraction-inversion pipeline is effective for a wide range of \textit{Mel-bins} possibly due to the distortion introduced by the phase estimation step during the inversion stage. For the \textit{Filtering} transform we do not perform a hyper-parameter search and use the transformation parameters described in Section~\ref{sec:filtering}.

\subsection{Detection Scores}
\label{sec:score}
Based on the above described search, we find the optimal hyper-parameters for each of the transforms and report the detection scores against all the attacks in Table~\ref{tab:nonadaptiveresult}. 
We observe that at optimal compression levels, all the input transforms listed in Section~\ref{sec:transformations} can achieve high discriminative performance against adversarial examples. As compared to targeted adversarial examples, it is harder to detect examples with universal adversarial perturbations. This is because universal perturbations attempt to distort the original transcription rather than targeting a very different phrase.
Interestingly, we find that the defense is effective even against the \textit{Qin-R} attack which incorporates noise simulation during training and leads to adversarial examples that are robust to small changes. We elaborate on this result in the following Section.

\subsection{Analysis of undefended and defended transcriptions}
\label{sec:transcriptevals}

\begin{figure}[H]
    \centering
    \includegraphics[width=0.95\columnwidth]{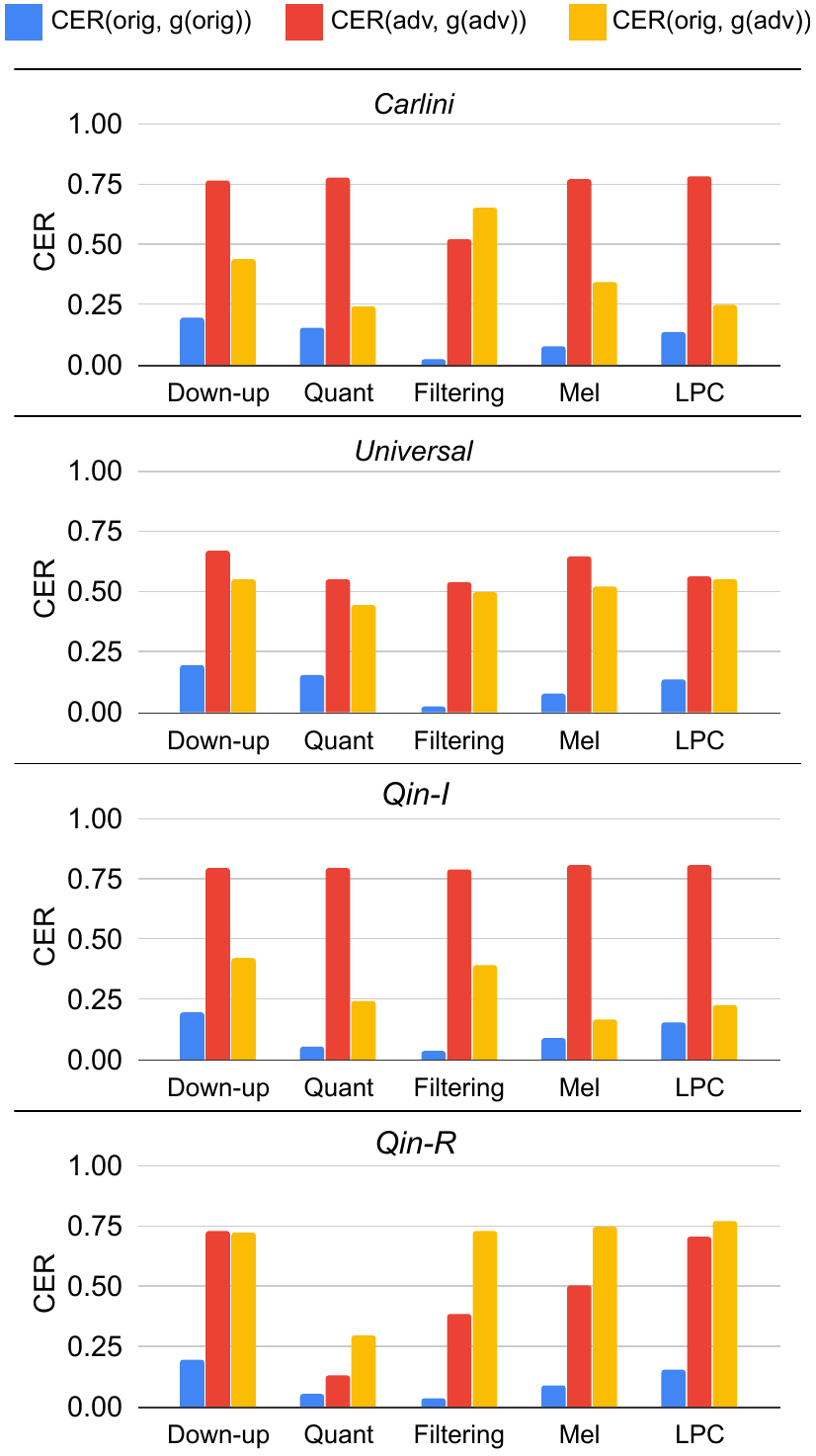}
    
    \caption{Mean Character Error Rate (CER) between the ASR transcriptions of un-transformed ($x$) and transformed ($g(x)$) audio. \textit{CER(orig,g(orig))} and \textit{CER(adv, g(adv))} indicate the CER between transcriptions of $x$ and $g(x)$ for benign and adversarial samples respectively. \textit{CER(orig, g(adv))} is the CER between the defended adversarial signal and its benign counterpart.} 
    \label{fig:cer_graphs}
\end{figure}

In Figure~\ref{fig:cer_graphs} we provide comparisons of Mean CER between transcriptions of audio before and after passing through a given transformation function ($g$) for both benign (\textit{orig}) and adversarial examples (\textit{adv}). Additionally, we also calculate the CER between the transcriptions of the defended adversarial example and its benign counterpart: $\textit{CER}(\textit{orig}, g(\textit{adv}))$.

The discriminative power of the detector is indicated by the difference between $\textit{CER}(\textit{orig}, g(\textit{orig}))$ (blue) and $\textit{CER}(\textit{adv}, g(\textit{adv}))$ (red). A high difference between the red and blue bar graphs in Figure~\ref{fig:cer_graphs} indicates easier detection of adversarial examples. From these results we can observe that detecting the \textit{Qin-I} attack is easier than detecting the \textit{Carlini}~\cite{targetattacks} attack. We can further deduce that detecting \textit{Universal} attacks is generally more difficult for any given transformation function compared to the \textit{Carlini} and \textit{Qin-I} attacks. 

The metric $\textit{CER}(\textit{orig}, g(\textit{adv}))$ helps evaluate the ability of the transformation function to recover the original transcript from the adversarial audio. A low $\textit{CER}(\textit{orig}, g(\textit{adv}))$ indicates better recovery of the original transcript. We find that for the imperceptible attack \textit{Qin-I}, the recovery rate of the original transcript is higher than any other attack indicating that the adversarial perturbation is 
unstable to small changes in inputs. 

The \textit{Qin-R} attack has a lower $\textit{CER}(\textit{adv}, g(\textit{adv}))$ for most transformations as compared to \textit{Qin-I} which suggests that the adversarial perturbation generated by the \textit{Qin-R} attack is relatively more robust to input transformations. Also, recovering the original transcription is much harder as compared to \textit{Qin-I} and is indicated by higher $\textit{CER}(\textit{orig}, g(\textit{adv}))$ values. However, there is still a significant difference between the blue and red bar graphs for \textit{Qin-R}, which can be used to discriminate between adversarial and benign samples. This result is consistent with the high detection accuracy reported in Table~\ref{tab:nonadaptiveresult}, since the transformations are successful in disrupting the adversarial perturbations.

\begin{figure*}[h]
    \centering
    \includegraphics[width=1.0\textwidth]{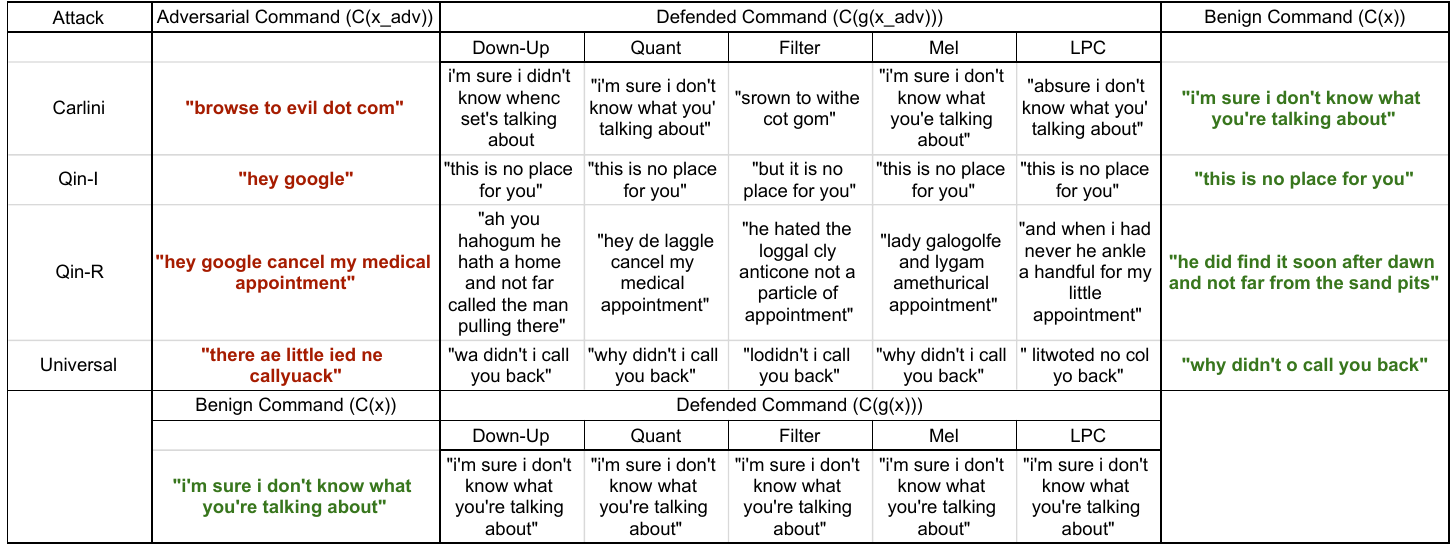}
    
    \caption{Sample transcriptions of un-transformed($x$) and transformed audio($g(x)$) for both benign and adversarial examples.}
    \label{fig:transcript}
\end{figure*}

We provide a few sample transcriptions from our experiments in Figure~\ref{fig:transcript}. 
The green commands indicate the transcriptions from benign audio samples, while the red transcriptions refer to adversarial commands from each attack type.  
Overall, the results in Figure~\ref{fig:cer_graphs} and Figure~\ref{fig:transcript} demonstrate that the ability to recover benign commands is dependent on the type of attack and varies for each input transformation function.

\subsection{Timing analysis}

To implement our defense framework in practice, we have to perform two forward passes through our ASR model to obtain the transcriptions $C(x)$ and $C(g(x))$. It is ideal to parallelize these two forward passes, so that the only computational overhead introduced by the defense is that of the transformation function $g$. Table~\ref{tab:timing} provides the average Wall-Clock time in seconds of each transformation function averaged over the 100 audio files (entire test set). 
Since some of our transformation functions were implemented solely on CPU, we provide timing comparisons for all implementations on the Intel Xeon CPU platform.
The average inference time over the test set for Mozilla Deepspeech ASR model is 2.540 seconds and that of Google Lingvo ASR model is 4.212 seconds on the Intel Xeon CPU Platform.

\begin{table}[htp]
\centering
\begin{tabular}{lc}
\toprule
Process & Avg. Wall-Clock time (s)\\ 
\midrule
Deepspeech ASR & 2.540\\
Lingvo ASR & 4.212\\
\midrule
Downsampling-Upsampling & 0.148 \\
Quantization-Dequantization & 0.001  \\
Filtering & 0.035 \\
Mel Extraction - Inversion & 0.569 \\
LPC & 0.781  \\
\bottomrule
\end{tabular}
\caption{Average Wall-Clock time in seconds required for transcription of audio by ASR models and each transformation function on Intel Xeon CPU platform. The Wall-Clock time is averaged over the entire test set.}
\label{tab:timing}
\end{table}

\section{Adaptive Attack}
\label{sec:adaptivemethodology}

While our defense framework can accurately discriminate adversarial from benign examples
for existing attacks, it only offers security in a 
``zero-knowledge''
attack scenario where the attacker is not aware of the defense being present. 
As motivated in Section~\ref{sec:related_adaptive}, in order to use our defense framework reliably in practice, it is important to evaluate it against an adaptive adversary who has complete knowledge of the defense and intend to design a perturbation that can bypass the defense mechanism.

In the adaptive attack setting, we will focus on the more impactful targeted attack scenario, where the adversary designs an adversarial perturbation that causes the victim ASR system to transcribe the input audio into a specific target phrase. 
In order to bypass the proposed defense framework, the adversary must craft an adversarial perturbation such that the transcription of $C(x_{\mathit{adv}})$ and $C(g(x_{\mathit{adv}}))$ match closely with each other and the target transcription $\tau$. Therefore, to craft such a perturbation $\delta$, the adversary aims to optimize the following problem: 
\begin{align*}
  \textit{minimize: } \;& |\delta|_\infty + c_1 \cdot \ell(x+\delta, \tau) + c_2 \cdot \ell(g(x+\delta), \tau)
\end{align*}
where, $\ell(x', t) = \text{CTC-Loss}(C(x'), t)$ and $c_1$ and $c_2$ are hyper-parameters that control the weights of the respective loss terms. Since optimization process over the $L_\infty$ metric is often unstable~\cite{targetattacks}, we modify our optimization objective as follows:
\begin{equation}
\begin{split}
  \textit{minimize: } \;& c\cdot|\delta|^2_2 + c_1 \cdot \ell(x+\delta, \tau) + c_2 \cdot \ell(g(x+\delta), \tau)\\ 
  &\textit{such that\ } |\delta|_\infty < \epsilon
  \end{split}
  \label{eq:adaptive}
\end{equation}

\subsection{Gradient Estimation for Adaptive Attack}
\label{sec:gradientestimation}
To 
solve the optimization problem given by equation~\ref{eq:adaptive} using gradient descent, the attacker must back-propagate the CTC-Loss through the ASR model and the input transformation function $g$. In case a differentiable implementation of  $g$ is not available, we use the Backward Pass Differentiable Approximation (BPDA) technique~\cite{obfuscated-gradients} to craft adversarial examples. That is, during the forward pass we use the exact implementation of the transformation function as used in our defense framework. During the backward pass, we use an approximate gradient implementation of the transformation $g$. We first perform the adaptive attack using the straight-through gradient estimator~\cite{obfuscated-gradients}. That is, we assume that the gradient of the loss with respect to the input $x$ to be the same as the gradient of the loss with respect to $g(x)$:
\begin{equation}
\left. \nabla_x \ell(g(x)) \right|_{x = \hat{x}} \approx \left. \nabla_x \ell(x) \right|_{x = g(\hat{x})}.
\end{equation}

In our experiments, we find that the straight-through estimator is effective in breaking the Quantization-Dequantization and Filtering transformation functions at low perturbation levels. 
However, using a more accurate gradient estimate can lead to a stronger attack. Specifically for the Mel extraction-inversion and LPC transformations, we find that using a straight-through gradient estimator does not work for solving the above optimization problem (Equation~\ref{eq:adaptive}). We discuss our results of using a straight-through gradient estimator for LPC transform in Appendix~\ref{sec:attackvarLPC}.
Also, using a straight-through estimator for the Downsampling-Upsampling transform results in high distortion for adversarial perturbations.
Therefore, we implement differentiable computational graphs for the following three transforms in TensorFlow:

\noindent \textbf{Downsampling-Upsampling:} We use TensorFlow's bi-linear resizing methods to first downsample the audio to the required sampling rate and then re-estimate the signal using bi-linear interpolation.

\noindent \textbf{Mel Extraction - Inversion:} For the Mel extraction-inversion transform we use TensorFlow's STFT implementation to obtain the magnitude spectrogram, then perform the Mel transform using matrix multiplication with the Mel basis, and estimate the waveform using the iterative Griffin-Lim~\cite{griffinlim} algorithm implemented in TensorFlow~\cite{githubgl}.

\noindent \textbf{LPC transform:} We implement the LPC analysis and synthesis process in TensorFlow. Specifically, for each window in the original waveform, we first estimate LPC coefficients by solving the linear regression problem given by Equation~\ref{eq:lpclinearregression}. Next, for the reconstruction process, we generate the residual excitation signal using the exact same implementation as used in our defense. We also fix the random seed of the excitation generator in both our defense and our adaptive attacks for a complete knowledge white box attack scenario. Next, we implement auto-regressive filtering of the residual signal with the LPC coefficients for that window to synthesize the signal for the given window. Finally, we add and combine the filtered signal for each overlapping window to generate the transformed audio.

Note that for all the adaptive attacks, we use the original defense implementations in the forward pass and use the differentiable implementation only during the backward pass. 

\subsection{Adaptive Attack Algorithm}
Algorithm \ref{algorithm1} details our adaptive attack implementation. We closely follow the targeted attack implementation in~\cite{targetattacks} and incorporate the optimization objective of our adaptive attack specified by Equation~\ref{eq:adaptive} and BPDA. We choose $c_1$ = $c_2$ = 1 since both loss terms have the same order of magnitude. Following the default open source implementation of~\cite{targetattacks}, we do not penalize $L_{2}$ distortion. 
We optimize for 5000 iterations and use a learning rate of 10. Any time the attack succeeds, we re-scale the perturbation bound by a factor of 0.8 to encourage less distorted (quieter) adversarial examples. 

\begin{algorithm}[h]
\caption{Adaptive attack algorithm}\label{universaltrain}
\begin{algorithmic}[1]
\State Initialize $\textit{rescaleFactor} \gets 1$
\State Initialize $\delta \gets 0$
\State Initialize $\textit{bestDelta} \gets \textit{null}$
\For { \textit{iterNum} in 1 to \textit{MaxIters}}
\State {$\textit{loss} \gets c \cdot |\delta|^2_2 + c_1 \cdot \ell(x+\delta, t) + c_2 \cdot \ell(g(x+\delta), t) $}
\State {$\nabla \delta \gets \textit{BPDA}(\textit{loss}, \delta)$}
\State{$\delta \gets \delta - \alpha \textit{ sign} (\nabla \delta)$}
\State {$\delta \gets \textit{rescaleFactor} * \textit{clip}_\epsilon (\delta)$}
\If {$C(x + \delta) = C( g(x + \delta)) = \tau$}
\State {$\textit{bestDelta} \gets \delta$}
\State {$\textit{rescaleFactor} \gets \textit{rescaleFactor} \times 0.8 $}
\EndIf

\EndFor
\If{\textit{bestDelta} is \textit{null}}
\State {$\textit{bestDelta} \gets \delta$}
\EndIf
\State {return $(x + \textit{bestDelta})$}
\end{algorithmic}
\label{algorithm1}
\end{algorithm}

\section{Adaptive Attack Evaluation}
\label{sec:adaptiveresults}

\begin{table*}[h]
\centering
\resizebox{1.0\textwidth}{!}{%
\begin{tabular}{l|rrc|cccc|cc}
\multicolumn{1}{c}{} &
\multicolumn{3}{c}{\emph{Distortion metrics}} &
\multicolumn{4}{c}{\emph{Attack Performance}} &
\multicolumn{2}{c}{\emph{Detection Scores}}\\
\toprule
Defense & $\epsilon_\infty$ &  $|\delta|_\infty$ & $\textit{dB}_x(\delta)$ & SR ($x_{\textit{adv}}$) & SR ($g(x_{\textit{adv}})$)  & $\textit{CER}(x_{\textit{\textit{adv}}}, \tau)$ & $\textit{CER}(g(x_{\textit{adv}}), \tau)$ & AUC & Acc.\\
\midrule
None & 500 & 81 & -45.3  & 100\% & - & 0.00 & - & - & -\\
\midrule
Downsampling - Upsampling & 500 & \textbf{342} & -32.7  & 100\% & 78\% & 0.00 & 0.05 & 0.31 & 50.0\%\\
Quantization - Dequantization & 500 & \textbf{215} & -36.7 & 100\% & 81\% & 0.00 & 0.01 & 0.11 & 50.0\%\\
Filtering  & 500 & \textbf{92} & -44.1 & 91\% & 72\% & 0.01 & 0.02 & 0.45 & 50.0\%\\
Mel Extraction - Inversion & 500 & 500 & -29.4 & 34\% & 0\% & 0.11 & 0.44 & 0.97 & 95.5\%\\
LPC & 500 & 500 & -29.4 & 43\% & 0\% & 0.06 & 0.51 & 0.94 & 86.0\%\\
\midrule
Mel Extraction - Inversion & 1000 & 1000 & -23.5 & 53\% & 0\% & 0.05 & 0.34 & 0.92 & 84.0\%\\

LPC & 1000 & 1000 & -23.5 & 72\% & 0\% & 0.01 & 0.29 & 0.77 & 72.5\%\\

\midrule
Mel Extraction - Inversion & 4000 & \textbf{2461} & -15.1 & 100\% & 31\% & 0.00 & 0.08 & 0.48 & 50.0\%\\
LPC & 4000 & \textbf{2167} & -16.7 & 100\% & 73\% & 0.0 & 0.03 & 0.21 & 50.0\%\\
\bottomrule
\end{tabular}
}
\caption{Adaptive attack evaluations against different transformation functions. $\epsilon_\infty$ is the initial $L_\infty$ bound used in the attack algorithm and $\delta_\infty$ is the mean $L_\infty$ norm of the perturbations obtained after applying the adaptive attack algorithm. Bolded values indicate the $\delta_\infty$ required to completely break (AUC $\leq$ 0.5) a particular transformation function based defense.
$\textit{dB}_x(\delta)$ is the relative loudness of the perturbation with respect to the examples in the dataset (the lower the quieter).
SR ($x_{\mathit{adv}}$) and SR $(g(x_{\mathit{adv}}))$ indicate the attack success rate for un-transformed ($x_{\mathit{adv}}$) and transformed audio $(g(x_{\mathit{adv}}))$ respectively obtained using the adaptive attack algorithm on a given transformation function.
}
\label{tab:adaptive}
\end{table*}

In this section, we test the limits of our defense and evaluate the breaking point for each transformation function through adaptive attacks in white box setting.
We conduct adaptive attack evaluations on the same dataset used in our previous experiments. The victim ASR for the adaptive attack is the Mozilla DeepSpeech model. In order to evaluate the imperceptibility of adversarial perturbations, we quantify the distortion of adversarial perturbations as follows.\\

\noindent \textbf{Distortion Metrics and Relative Loudness:} We first implement adaptive attacks using an initial distortion bound $|\epsilon|_\infty = 500$. Note that we are using a 16-bit waveform representation which means that the waveform samples are in the range -32768 to 32768. An $L_\infty$ distortion of 500 is fairly perceptible although it does not completely mask the original signal.\footnote{ Audio Examples: \url{https://waveguard.herokuapp.com} } Along with the $L_\infty$ norm of the perturbation, we report another related metric  $\textit{dB}_x(\delta)$~\cite{targetattacks,neekhara2019universal} that measures the relative loudness of the perturbation with respect to the original signal in Decibels(dB). The metric $\textit{dB}_x(\delta)$ is defined as follows:
\begin{equation}
\begin{split}
  & \mathit{dB}(x) = \mathit{max}_i 20 \log_{10}(x_i)\\ 
  &\mathit{dB}_x(\delta) = \mathit{dB}(\delta) - \mathit{dB}(x)
  \end{split}
  \label{eq:db}
\end{equation}
The more negative $\textit{dB}_x(\delta)$ is, the quieter is the adversarial perturbation. For comparison, -31 dB is roughly the difference between ambient noise in a quiet room and a person talking~\cite{targetattacks}. While we start with an initial $L_\infty$ ($\epsilon_\infty$) bound of 500 in our experiments, the final distortion norm ($\delta_\infty$) can be much smaller than the initial bound. This is because our optimization objective penalizes high distortion amounts and our algorithm re-scales the perturbation bound by a factor of 0.8 every time the attack succeeds.

Generally, prior work on attacks to ASR systems apply particular attention to minimize perturbation distortions, in order to encourage imperceptibility of adversarial audio. Towards this goal of generating imperceptible adversarial examples, Qin et al.~\cite{qin2019imperceptible} and Universal~\cite{neekhara2019universal} generate examples with maximum allowed distortion of $L_\infty$ = 400, while Carlini et al.~\cite{targetattacks} generate examples with maximum distortion of $L_\infty$ = 100. However for conducting our adaptive attack evaluation, since we aim to test the breaking point of each transformation function, we generate adversarial perturbations at much higher $L_\infty$ bounds (500, 1000, 4000) that are significantly more audible to the human ear. 

\begin{figure}[htbp]
\centering
\subfloat[Downsampling-upsampling, Quantization and Filtering]{%
  \includegraphics[clip,width=1.0\columnwidth]{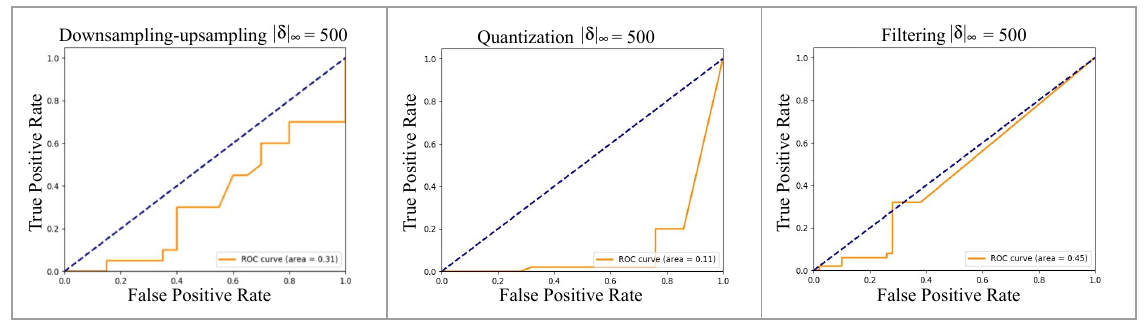}%
}

\subfloat[Linear Predictive Coding (LPC)]{%
  \includegraphics[clip,width=1.0\columnwidth]{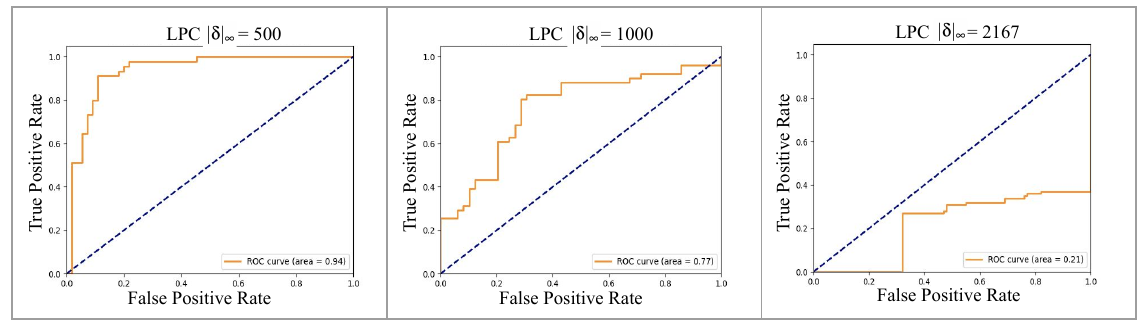}%
}

\subfloat[Mel Extraction - Inversion]{%
  \includegraphics[clip,width=1.0\columnwidth]{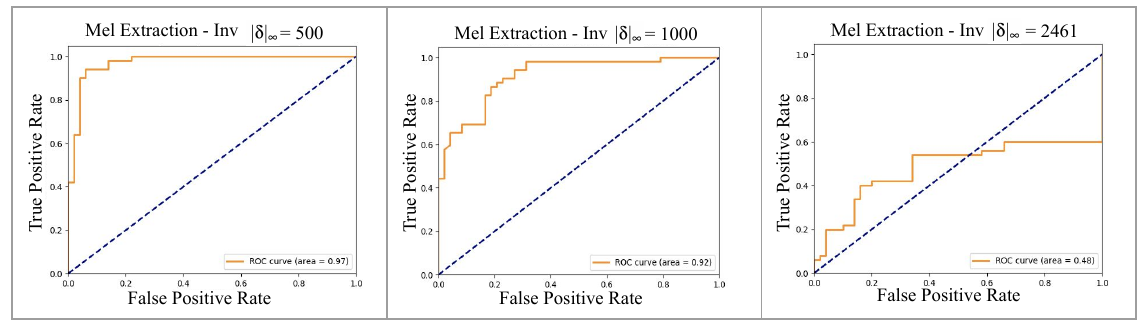}%
}

\caption{Detection ROC curves for different transformation functions against adaptive attacks (Section~\ref{sec:adaptiveresults}) with various magnitudes of adversarial perturbation ($|\delta|_\infty$).}
\label{fig:ROC_adaptive}
\end{figure}

Table~\ref{tab:adaptive} presents the results for our adaptive attack against various input transformation functions. 
We provide the Receiver Operating Characteristic (ROC) of the detector in the adaptive attack settings for different transformation functions under different magnitudes of perturbation in Figure~\ref{fig:ROC_adaptive}. A \textit{true positive} implies an example that is adversarial and is correctly identified as adversarial.
We evaluate the adaptive attacks on two aspects: 1) \textit{Attack Performance:} How successful was the adaptive attack in its objective? 2) \textit{Detection Scores:} How effective is our detector for the adversarial audios generated by the attack?

For the adaptive attacks against the \textit{Downsampling-upsampling}, \textit{Quantization-Dequantization} and \textit{Filtering} transforms, we achieve low CER between the target transcription and transcriptions for $x_{\textit{adv}}$ and $g(x_{\textit{adv}})$ ($\textit{CER}(x_{\textit{adv}}, \tau)$ and $\textit{CER}(g(x_{\textit{adv}}))$ respectively). 
This makes it harder for the detector to discriminate between adversarial and benign samples thereby resulting in a drastic drop in detector AUC and accuracy scores as compared to the non-adaptive scenario. Amongst these three transformations, bypassing \textit{Downsampling-upsampling} requires the highest amount of perturbation ($\delta_\infty = 342$) indicating that it serves as a more robust defense transformation as compared to \textit{Quantization-Dequantization} and \textit{Filtering}. The columns $SR (x_{\textit{adv}})$ and $SR (g(x_{\textit{adv}}))$ indicate the percentage of examples that transcribed exactly to the target phrase for the un-transformed and transformed adversarial inputs respectively. 

The calibration of the detection threshold depends on the use case of the ASR system---for a user facing ASR system, the number of legitimate commands would usually be very high as compared to the number of adversarial commands. 
Therefore, the false positive rate needs to be extremely low for such ASR systems. 
As shown in Figure~\ref{fig:ROC} ( Appendix~\ref{sec:roc_non_adaptive}), in the non-adaptive attack scenario, we are able to achieve a  very high true positive rate at 0\% false positive rate for the targeted adversarial attacks (Carlini and Qin-I) for all transformation functions. Therefore a low detection threshold can be reliable against non-adaptive adversaries and also not interfere with the user experience. In the adaptive attack scenario, while both LPC and Mel inversion achieve higher AUC scores as compared to other transforms, Mel inversion transform gives the highest true positive rate at extremely low false positive rates. Therefore, amongst the transformation functions studied in our work, Mel Extraction and Inversion serves as the best defense choice for user facing ASR systems. 
\\

\noindent \textbf{Robustness of perceptually informed representations:}
For both Mel extraction-inversion and LPC transformations, although we observe a drop in the detector scores as compared to the non-adaptive attack setting, we are not able to completely bypass the defense using the initial distortion bound $\epsilon_\infty=500$. Note that a perturbation higher than this magnitude, has $\textit{dB}_x(\delta) > -29$ which is more audible than ambient noise in a quiet room ($\textit{dB}_x(\delta)=-31$)~\cite{dbnoise,carlini2017adversarial}. 
In order to test the limit at which the defense breaks, we successively increase the allowed magnitude of perturbation. We are able to completely break the defense (AUC $\leq$ 0.5) at $\delta_\infty=2479$ and  $\delta_\infty=2167$ for Mel extraction-inversion and LPC transforms respectively. These perturbations are more than $6\times$ higher than that required to break any of the other transformation functions studied in our work and more than $25\times$ higher than that required to fool an undefended model. This suggests that using perceptually informed intermediate representations prove to be more robust against adaptive attacks as compared to naive compression and decompression techniques.

\begin{figure}[htbp]
    \centering
    \includegraphics[width=1.0\columnwidth]{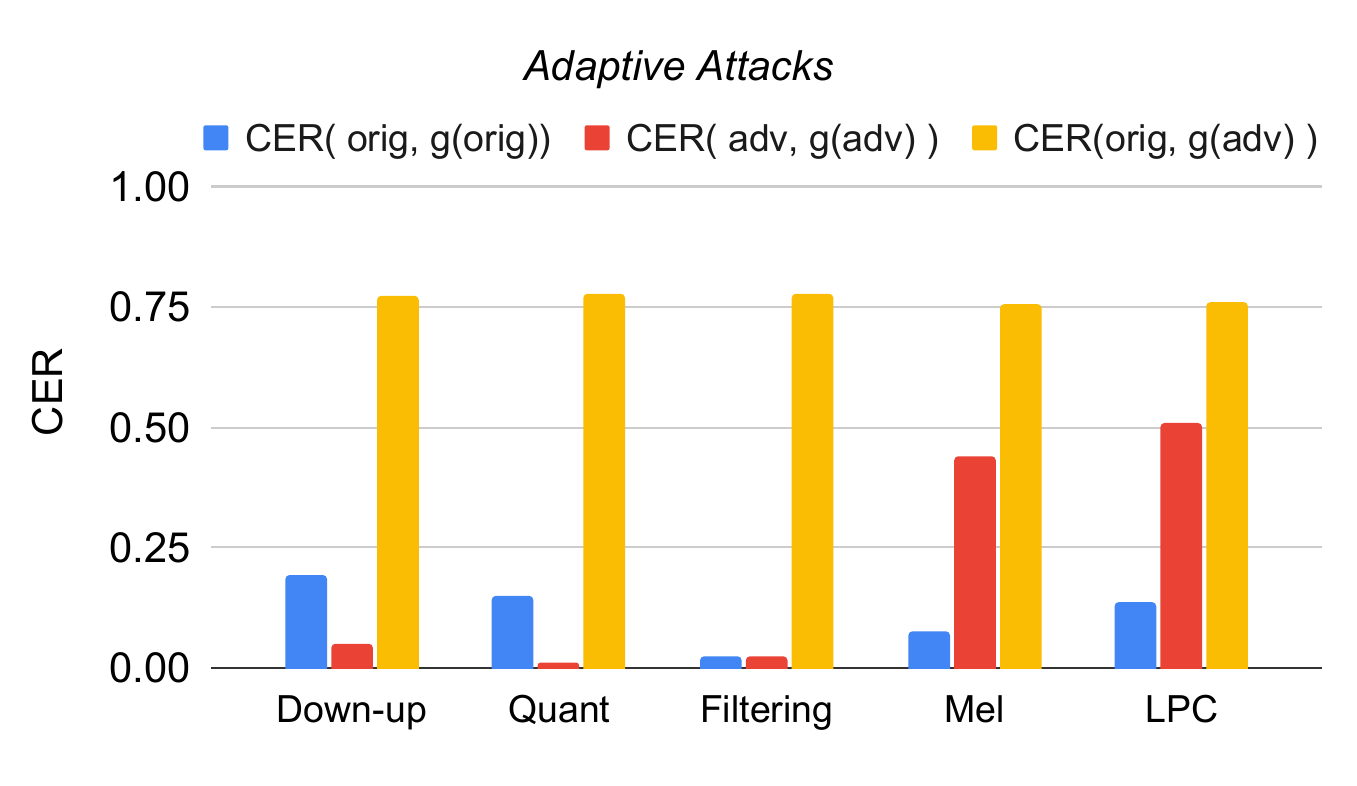}
    
    \caption{Mean CER between the ASR transcriptions of un-transformed ($x$) and transformed ($g(x)$) audio for adaptive attacks with an initial distortion $\epsilon_\infty = 500$. 
    } 
    \label{fig:adaptivecer}
\end{figure}

Figure~\ref{fig:adaptivecer} reports the same metrics as those reported in Figure~\ref{fig:cer_graphs} for the adaptive attack scenario with an initial $\epsilon_\infty=500$. The $CER(adv, g(adv))$ (red bar) drops below $CER(orig, g(orig))$ (blue bar) for \textit{Downsampling-upsampling}, \textit{Quantization-Dequantization} and \textit{Filtering} transforms thereby breaking these defenses. 
In contrast, the red bar for \textit{Mel extraction-inversion} and \textit{LPC} based defense is much higher than the blue bar indicating that the defense is more robust under this adaptive attack setting.

\section{Discussion}

\textbf{Do learnings from adversarial defenses in the image domain transfer over to the audio domain?}
We find that not all learnings about input-transformation based defenses in the image domain transfer to the speech recognition domain. It has been shown that input-transformation based adversarial defenses can be easily bypassed using robust or adaptive attacks for image classification systems~\cite{obfuscated-gradients,eot}. However, an ASR system is a substantially different architecture as compared to an image classification model. ASR systems operate on time-varying inputs and map each input frame to a language token. Since they rely on Recurrent Neural Networks (RNNs), the token prediction for each frame also depends on other frames in the signal. For targeted attacks, that are robust to a transformation $g$, we need to find an adversarial example $x_{\textit{audio}}$ such that both $x_{\textit{audio}}$ and $g(x_{\textit{audio}})$ map to the target language tokens across all time-steps. On the other hand, for the image classification problem, the adaptive attack goal is simpler: Find an image $x_{\textit{image}}$, such that both $x_{\textit{image}}$ and $g(x_{\textit{image}})$ map to the same class label. Therefore, in our adaptive attack experiments, we need to add significant amount of perturbation to bypass the defense even for simple transformation functions. We also find that adversarial attacks targeting undefended ASR models do not transfer to defended models even at high perturbation levels, in contrast to results reported in the image domain~\cite{tramer2020adaptive}. Details of this experiment are provided in Appendix~\ref{sec:transferattackappendix}. 

\section{Conclusion}
We present \textit{WaveGuard}, a framework for detecting audio adversarial inputs, to address the security threat faced by ASR systems. Our framework incorporates audio transformation functions and analyzes the ASR transcriptions of the original and transformed audio to detect adversarial inputs. We demonstrate that WaveGuard can reliably detect adversarial inputs from recently proposed and highly successful targeted and untargeted attacks on ASR systems. Furthermore, we evaluate WaveGuard in the presence of an \textit{adaptive} adversary who has complete knowledge of our defense. 
We find that only at significantly higher magnitudes of adversarial perturbation, which are audible to the human ear, can an adaptive adversary bypass transformations that compress input to perceptually informed audio representations. In contrast, naive audio transformation functions can be easily bypassed by an adaptive adversary using small inaudible amounts of perturbations.
This makes transformations such as LPC and Mel extraction-inversion more robust candidates for defense against audio adversarial attacks.

\section*{Acknowledgements}
We thank our reviewers for their valuable and comprehensive feedback. This work was supported by SRC under Task ID: 2899.001 and ARO under award number
W911NF-19-1-0317. 


\bibliography{example_paper}
\bibliographystyle{IEEEtran}
\clearpage
\section{Appendix}
\subsubsection{Receiver Operating Characteristic curves for Detection under Non-Adaptive Attacks}
\label{sec:roc_non_adaptive}
\begin{figure}[htp]
\centering
\subfloat[Downsampling-upsampling]{%
  \includegraphics[clip,width=1.0\columnwidth]{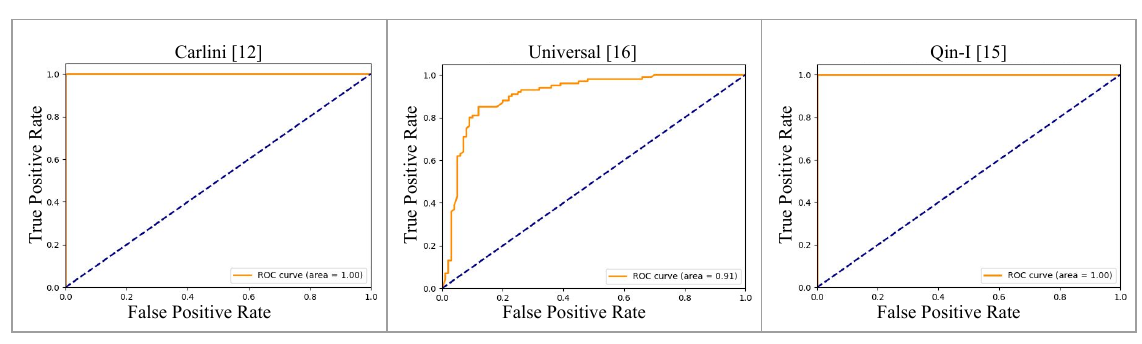}%
}

\subfloat[Quantization]{%
  \includegraphics[clip,width=1.0\columnwidth]{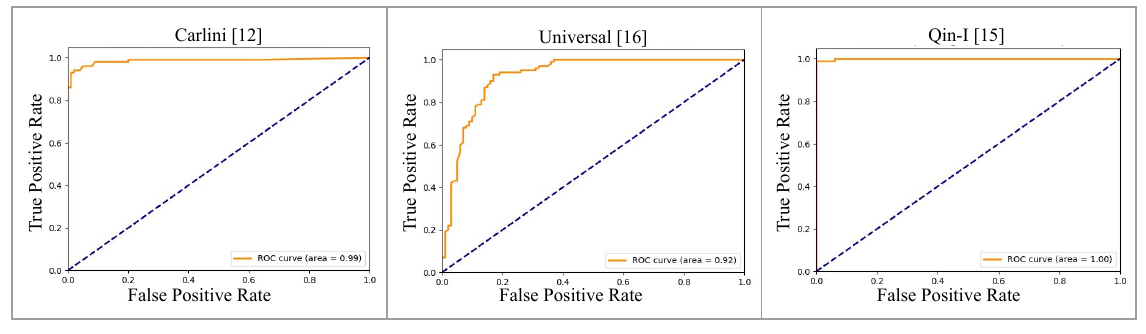}%
}

\subfloat[Filtering]{%
  \includegraphics[clip,width=1.0\columnwidth]{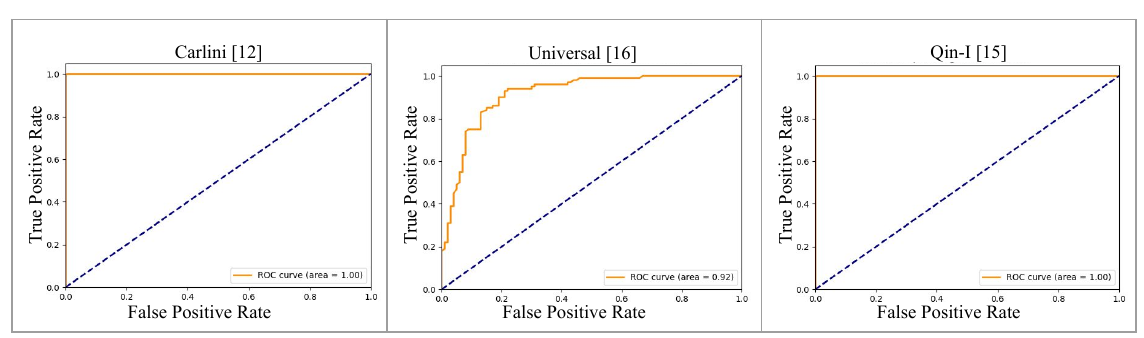}%
}

\subfloat[Linear Predictive Coding (LPC) ]{%
  \includegraphics[clip,width=1.0\columnwidth]{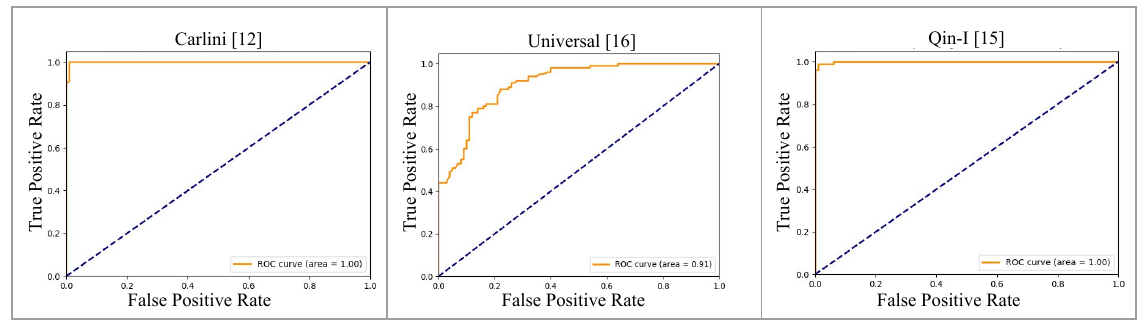}%
}

\subfloat[Mel Extraction - Inversion]{%
  \includegraphics[clip,width=1.0\columnwidth]{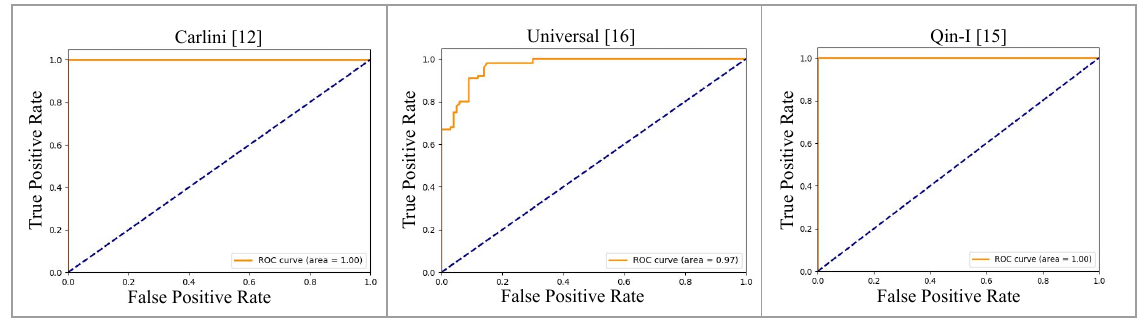}%
}

\caption{Detection ROC curves for different transformation functions against three attacks (Carlini~\cite{targetattacks}, Universal~\cite{neekhara2019universal}, Qin-I~\cite{qin2019imperceptible}) in the non-adaptive attack setting.}
\label{fig:ROC}
\end{figure}

We provide the Receiver Operating Characteristic (ROC) curves for our detection of non-adaptive adversarial attacks using various transformation functions against three different adversarial attacks in Figure~\ref{fig:ROC}. The AUC scores are reported in Table~\ref{tab:nonadaptiveresult} in Section~\ref{sec:score} and included with each of the plots below. A \textit{true positive} implies an example that is adversarial and is correctly identified as adversarial.

\subsubsection{Thresholds for Detection Accuracy}

Table~\ref{tab:threshold} lists the detection thresholds ($t)$ for various transformation functions for the two ASR systems studied in our work. We choose 50 original examples (separate from the first 100 used for evaluation) and construct 50 adversarial examples using each of the attack. This results in 100 adversarial-benign example pairs for DeepSpeech (constructed using Carlini~\cite{targetattacks} and Universal~\cite{neekhara2019universal} attacks) and 100 adversarial-benign example pairs for Google Lingvo (constructed using Qin-I and Qin-R attacks~\cite{qin2019imperceptible}). Using this dataset, we obtain the threshold that achieves the best detection accuracy for each defense separately for the two ASRs. The AUC metric is threshold independent. We do not change the threshold for adaptive attack evaluation and use the same threshold as listed in Table~\ref{tab:threshold}.

\begin{table}[h]
\centering
\resizebox{0.9\columnwidth}{!}{%
\begin{tabular}{@{}lcc@{}}
\toprule
Defense & \begin{tabular}[c]{@{}c@{}}Threshold - \\ DeepSpeech\end{tabular} & \begin{tabular}[c]{@{}c@{}}Threshold - \\ Lingvo\end{tabular} \\ \midrule
Downsampling - Upsampling & 0.48 & 0.48 \\
Quantization - Dequantization & 0.44 & 0.26 \\
Filtering & 0.32 & 0.31 \\
Mel Extraction - Inversion & 0.33 & 0.31 \\
LPC & 0.38 & 0.46 \\ \bottomrule
\end{tabular}
}
\caption{Detection Threshold when using each transformation function in WaveGuard framework for DeepSpeech and Lingvo ASR systems.}
\label{tab:threshold}
\end{table}

\subsubsection{Transfer Attacks from an Undefended Model}
\label{sec:transferattackappendix}
\begin{table}[h]
\setlength\tabcolsep{2pt}
\centering
\resizebox{1.0\columnwidth}{!}{%
\begin{tabular}{l|rc|cc|cc}
\multicolumn{1}{c}{} &
\multicolumn{2}{c}{\emph{Distortion metrics}} &
\multicolumn{2}{c}{\emph{Attack Performance}} &
\multicolumn{2}{c}{\emph{Detection Scores}}\\
\toprule
Defense & $|\delta|_\infty$ & $\textit{dB}_x(\delta)$ & $\textit{CER}(x_{\textit{\textit{adv}}}, \tau)$ & $\textit{CER}(g(x_{\textit{adv}}), \tau)$ & AUC & Acc.\\
\midrule
LPC  & 1000 & -23.5 & 0.0 & 0.80 & 0.99 & 98.5\%\\
LPC  & 2000 & -17.4 & 0.0 & 0.83 & 0.99 & 99.0\%\\
LPC  & 4000 & -11.4 & 0.0 & 0.81 & 0.99 & 97.0\%\\
LPC  & 8000 & -5.4 & 0.0 & 0.91 & 0.99 & 99.0\%\\
\midrule
Mel Ext - Inv  & 1000 & -23.5 & 0.0 & 0.81 & 0.99 & 98.5\%\\
Mel Ext - Inv  & 2000 & -17.4 & 0.0 & 0.88 & 0.99 & 97.5\%\\
Mel Ext - Inv  & 4000 & -11.4 & 0.0 & 0.89 & 0.99 & 98.0\%\\
Mel Ext - Inv  & 8000 & -5.4 & 0.0 & 0.92 & 0.99 & 98.5\%\\
\bottomrule
\end{tabular}
}
\caption{Evaluation of Mel Extraction - Inversion and LPC transform defense against perturbations targeting an undefended DeepSpeech ASR model at different levels of magnitude. 
}
\label{tab:transferlpc}
\end{table}

We additionally evaluate the robustness of Mel extraction-inversion and LPC transformations against transfer attacks from an undefended model. We craft targeted adversarial examples using~\cite{targetattacks} for DeepSpeech ASR at different perturbation levels by linearly scaling the perturbation to have the desired $L_\infty$ norm. Table~\ref{tab:transferlpc} shows the evaluations of transfer attack at different perturbation levels. We find that attacks targeting undefended models do not break the defense using these two transformation functions even at high perturbation levels. This is because the transcription of $g(x_{adv})$ is significantly different from the target transcription and transcription of $x_{adv}$ even at high perturbation levels thereby allowing our detector to consistently detect the adversarial samples.

\subsubsection{Straight-through Gradient Estimator for LPC}
\label{sec:attackvarLPC}

We find that the LPC transform cannot be broken in an adaptive attack scenario using BPDA attack with a straight-through gradient estimator (i.e assuming identity function as the gradient of transformation function $g$ during the backward pass). In our experiments, we started with an initial $\epsilon_\infty$ of 2000, and increased the initial distortion bound to 16000 but did not observe any improvement in the attack performance as the detector was still able to identify adversarial audio with 100\% accuracy. Therefore, using our BPDA attack algorithm, we do not arrive at a solution in which both $x$ and $g(x)$ transcribe to the target phrase even with a high amount of allowed distortion. This motivated us to design stronger adaptive attacks with differentiable LPC (Section~\ref{sec:gradientestimation}) to find distortion bounds over which LPC transforms are not able to reliably detect adversarial examples.

\begin{table}[h]
\setlength\tabcolsep{2pt}
\centering
\resizebox{1.0\columnwidth}{!}{%
\begin{tabular}{l|rrc|cc|cc}
\multicolumn{1}{c}{} &
\multicolumn{3}{c}{\emph{Distortion metrics}} &
\multicolumn{2}{c}{\emph{Attack Performance}} &
\multicolumn{2}{c}{\emph{Detection Scores}}\\
\toprule
Defense & $\epsilon_\infty$ &  $|\delta|_\infty$ & $\textit{dB}_x(\delta)$ & $\textit{CER}(x_{\textit{\textit{adv}}}, \tau)$ & $\textit{CER}(g(x_{\textit{adv}}), \tau)$ & AUC & Acc.\\
\midrule
LPC & 2000 & 2000 & -15.9 & 0.31 & 0.85 & 1.0 & 100\%\\
LPC & 16000 & 16000 & 2.1 & 0.34 & 0.85 & 1.0 & 100\%\\
\bottomrule
\end{tabular}
}
\caption{Evaluation of LPC transform against straight-through gradient estimator.
}
\label{tab:adaptivelpc}
\end{table}

\clearpage







\end{document}